\documentclass[aps,pre,amsmath,amsfonts,onecolumn,longbibliography,superscriptaddress,floatfix]{revtex4-1}

\usepackage{amsmath}

\usepackage{caption}
\usepackage{subcaption}

\usepackage{graphicx}

\begin{document}

\title{Gnomonious Projections for Bend-Free Textures:  Thoughts on the Splay-Twist Phase}

\author{N. Chaturvedi}
\author{Randall D. Kamien}

\affiliation{Department of Physics and Astronomy, University of Pennsylvania, Philadelphia,
PA, 19104-6396, USA}




\begin{abstract}
The Hopf fibration has inspired any number of geometric structures in physical systems, in particular in chiral liquid crystalline materials.  Because the Hopf fibration lives on the three sphere, $\mathbb{S}^3$, some method of projection or distortion must be employed to realize textures in flat space.  
Here, we explore the geodesic-preserving gnomonic projection of the Hopf fibration, and show that this could be the basis for a new liquid crystalline texture with only splay and twist. We outline the structure and show that it is defined by the tangent vectors along the straight line rulings on a series of hyperboloids. The phase is defined by a lack of bend deformations in the texture, and is reminiscent of the splay-bend and twist-bend nematic phases. We show that domains of this phase may be stabilized through anchoring and saddle-splay.  
\end{abstract}


\maketitle
\section{Introduction}\label{intro}

Topology and geometry have played an important role in illuminating the myriad textures that liquid crystals can make \cite{Frank,klemanbook,bouligand,sadoc20093}. For instance, the blue phase of cholesteric liquid crystals has a frustrated texture in $\mathbb{R}^3$ with regions of double twist seperated by a lattice of disclination lines \cite{pansu1987textures, dandoloff1993twisted}. The texture can be made frustration-free when placed in the curved space of $\mathbb{S}^3$ \cite{sethna1983frustration}. Further, the smooth double twist structure in $\mathbb{S}^3$ is tangent to a fascinating fibration of $\mathbb{S}^3$ with great circles, known as the Hopf fibration \cite{dandoloff1987blue, mosseri2012hopf}. In this paper, we take inspiration from this history to explore projections of the Hopf fibration for a twisted nematic texture that is bend-free. We show in section \ref{hopf} that in order to make a bend-free texture, we need a projection that preserves geodesics, the gnomonic projection. We call the texture which results from taking the gnomonic projection of the Hopf fibration `splay-twist'.

\maketitle
The splay-twist texture is reminiscent of the twist-bend and splay-bend phases that lack splay and twist, respectively \cite{dozov2001spontaneous, borshch2013nematic, henderson2011methylene}. Given the existence of these two phases, it is not unreasonable to expect the existence of the third nematic phase. In the last few decades, the modulated nematic phase known as twist-bend has received  attention due to its unique chiral symmetry breaking properties and giant flexoelectric constants both intrinsic \cite{harden2006giant, harden2010giant} and structurally enhanced \cite{mld,shamid2013statistical, prasang2008spontaneous, meyer2015temperature, cestari2011phase, chen2014twist}. Further, the splay nematic phase was recently discovered as the second modulated nematic phase in experiment \cite{mertelj2018splay}. Though liquid crystalline materials show a rich variety of structures and phases, only a few distinct nematic (chiral or not) phases have been discovered. 

We show that  `splay-twist' domains consists of molecules following a straight line ruling on the surface of a series of hyperboloids. The angle of the straight lines with the hyperboloid axis increases with increasing distance from the center of the texture, allowing the architecture to be space-filling. The splay-twist texture is shown on the left hand sides of Figs. \ref{fig:hyper} and \ref{fig:loops}. Further, at large distances from the center, the phase resembles a hedgehog nematic texture. In section \ref{geometry} we examine the elastic free energy of the splay-twist phase, and show that the phase can be stabilized with saddle-splay as in the blue-phase, or with anchoring conditions, like hedgehog nematic textures. 

\section{The Hopf Fibration and Its Gnonomic Projection}\label{hopf}

Cholesteric liquid crystal blue phases are frustrated systems and consist of lattices of double twist regions seperated by lattices of disclination lines. When the double twist structure is put in the curved space of $\mathbb{S}^3$, however, the system ceases to be frustrated, and its texture is space filling \cite{sethna1983frustration}. In $\mathbb{S}^3$, the nematic structure of the blue phase is tangent to a fibration of $\mathbb{S}^3$ with great circles, known as the Hopf fibration. Indeed, the imprint of the Hopf fibration can be seen in at least ``seven different physical systems'' \cite{urbantke2003hopf}, and has previously been studied in the context of nematic liquid crystal defects \cite{sadoc20093, chen2013generating}.

In the following, ${\bf V}=(X,Y,Z)=(\sin\theta_0\cos\phi_0,\sin\theta_0\sin\phi_0,\cos\theta_0)$ are coordinates on $\mathbb{S}^2$, $(\tilde x,\tilde y,\tilde z,\tilde w)$ are coordinates in $\mathbb{R}^4$ and $(x,y,z)$ are coordinates in $\mathbb{R}^3$.  Recall that the Hopf fibration is a map from $\mathbb{S}^3\rightarrow \mathbb{S}^2$.  We parameterize $\mathbb{S}^3\subset \mathbb{R}^4$  by $(\tilde{x}, \tilde{y}, \tilde{z}, \tilde{w})$ such that $\tilde{x}^2+\tilde{y}^2+\tilde{z}^2+(\tilde{w}-1)^2=1$ (a three-sphere in $\mathbb{R}^4$ sitting ``above'' $\mathbb{R}^3$). The preimage of the vector $\bf V$ is then a great circle on $\mathbb{S}^3$ parameterized by an angle $\psi\in[-\pi,\pi)$,

\begin{align}
\tilde{x} &=  (X\sin\psi-Y\cos\psi)/\sqrt{2(1+Z)}\nonumber\\
\tilde{y}&= (Y\sin\psi+X\cos\psi)/\sqrt{2(1+Z)}\nonumber\\
\tilde{z}&= \sqrt{(1+Z)/2} \cos\psi\nonumber\\
\tilde{w}&= \sqrt{(1+Z)/2} \sin\psi +1
\label{G}.
\end{align}
Recall that the gnomonic projection of $\mathbb{S}^2$ is a projection through the center of the sphere to the two-dimensional plane upon which the sphere sits.  It is similar to, but different from the stereographic projection which projects from the same sphere to the same plane but through the north pole.  In the latter, the entire sphere is conformally mapped to the Riemann sphere, $\mathbb{R}^2\cup\{\infty\}$, sending circles to circles.  In the gnomonic projection, however, only the southern hemisphere can be mapped to $\mathbb{R}^2$ -- the equator does not map to a unique point at infinity.  However, if we consider a great circle on $\mathbb{S}^2$ then it lies in a plane with the sphere center.  That plane intersects $\mathbb{R}^2$ in a line -- the gnomonic projection preserves geodesics!  Similarly, the gnomonic projection from $\mathbb{S}^3$ to $\mathbb{R}^3$ will map great circles into straight lines ({\sl i.e.}, the interesection of $\mathbb{R}^2$ with $\mathbb{R}^3$ in $\mathbb{R}^4$ is a straight line). The appearance of the gnomonic projection is not surprising then, since we are looking for a texture without bend, which implies the presence of straight lines. This suggests that we look at a projection of the Hopf fibration that maps great circles to straight lines or, in other words, preserves geodesics.  For a review of this and other projections of the sphere, see \cite{ng}.

The projection from $\mathbb{S}^3$ to $\mathbb{R}^3$ is
\begin{align}
(x,y,z) = (\tilde{x},\tilde{y},\tilde{z})/(1-\tilde{w})
\end{align}
This can only project one hemisphere of $\mathbb{S}^3$ to $\mathbb{R}^3$ since $(1-\tilde{w})$ vanishes on the equator of $\mathbb{S}^3$ where $\tilde{x}^2+\tilde{y}^2+\tilde{z}^2=1$.  Employing the fibration in (\ref{G}) we find
$1-\tilde{w} = \sqrt{{1\over 2}(1+Z)}\sin\psi$ and so
\begin{equation}
(x,y,z)_{\bf V} = \frac{1}{ (1+Z)}\big(X-Y\cot\psi, Y+X\cot\psi, (1+Z)\cot\psi\big)
\label{spatialgnomonic}
\end{equation}
For each ${\bf V}$ this is the equation for a straight line parameterized by $\cot\psi$ with $\psi\in[-\pi,0]$ (the southern hemisphere) or $s\equiv\cot\psi\in(-\infty,\infty)$. The tangent to each of these lines defines the nematic director:
\begin{align}
{\bf n} = [-\sin(\theta_0/2)\sin\phi_0, \sin(\theta_0/2)\cos\phi_0,\cos(\theta_0/2)]
\end{align}
Thus, the process of taking the preimage of the Hopf fibration and then a gnomonic projection amounts to a map that rotates the azimuthal angle by $\pi/2$ and halves the polar angle - a meron configuration on the two-sphere at infinity \cite{CDG}.   We can also think of the gnomonic projection as being generated by the method of characteristics.  We can merely project from $\mathbb{S}^3$ onto the $z=0$ plane of $\mathbb{R}^3$ and then generate the characterstics generated by $({\bf n}\cdot{\nabla}){\bf n}=0$.  In the last section we will discuss the conditions that prevent any shocks from occuring -- when two characterstics collide.  Note that although we draw the characteristics as lines, we are imagining here only a nematic texture made of short molecules.   Were we considering, instead, polymer nematics, then we would have to account for the increased $K_1$ associated with splay-density coupling \cite{meyerrb}.

Continuing, we write $\rho^2=x^2+y^2$, and find
\begin{equation}
\rho^2 = (1+z^2)\frac{X^2+Y^2}{(1+Z)^2} = (1+z^2) \tan^2(\theta_0/2)
\end{equation}
and so we see that for each $\theta_0$ the Hopf fibration sweeps out a hyperboloid with a waist radius of $\tan(\theta_0/2)$.  As $\theta_0$ grows from $0$ to $\pi$ the hyperboloids nest with polar angle $\theta_0/2$, eventually forming an azimuthal defect in $\bf n$ at infinite $\rho$, when $\theta_0=\pi$.  To find $\theta_0$ as a function of $(x,y,z)$, we invert (\ref{spatialgnomonic}) and find
\begin{eqnarray}
\frac{Y}{1+Z}&=&\frac{y-xz}{1+z^2}\nonumber\\
\frac{X}{1+Z}&=&\frac{x+yz}{1+z^2}
\end{eqnarray}
so that $\tan(\theta_0/2) = \sqrt{X^2+Y^2}/(1+Z)= \rho/\sqrt{1+z^2} $ and $\cos(\theta_0/2) = \sqrt{1+z^2}/\sqrt{1+r^2}$ where $r^2=x^2+y^2+z^2$.
Finally, we get the nematic director field as a function of $(x,y,z)$:
\begin{align}
{\bf n} = \left[\frac{-y+xz}{ \sqrt{1+z^2}\sqrt{1+r^2}} ,\frac{x+yz}{  \sqrt{1+z^2}\sqrt{1+r^2}}, \frac{\sqrt{1+z^2}}{\sqrt{1+r^2}}\right] \label{nematicdirector}
\end{align}
Note that as $r\rightarrow\infty$, ${\bf n} \rightarrow (x/r,y/r,z/r)$ in the upper half-space ($z>0$), as in a hedgehog configuration.  For $z<0$, however, ${\bf n} \rightarrow -(x/r,y/r,z/r)$.  As a result, there is no net hedgehog charge in any finite volume.  This is not a surprise since the core is defect-free.  The difference between this configuration and the true hedgehog can be seen at $z=0$.  When $z=0$, the texture is a meron configuration, with ${\bf n}=\hat z$ at the origin and a winding of $+1$ as $r\rightarrow\infty$ \cite{selinger,CDG}.  Were this texture a true hedgehog, there would be a {\sl radial} defect at infinity.  The winding near the $z=0$ plane is responsible for zeroing the hedgehog charge.  Finally, this configuration would be dilated or contracted had we projected from the three sphere of radius $\lambda\ne 1$.  In this case we would replace the $\mathbb{R}^3$ coordinates with $(x,y,z)\rightarrow (x,y,z)/\lambda$.  We will only focus on $\lambda=1$ in the following calculations since this just amounts to a change in overall scale.

\begin{figure}[t]
\centering
\includegraphics[width=\linewidth]{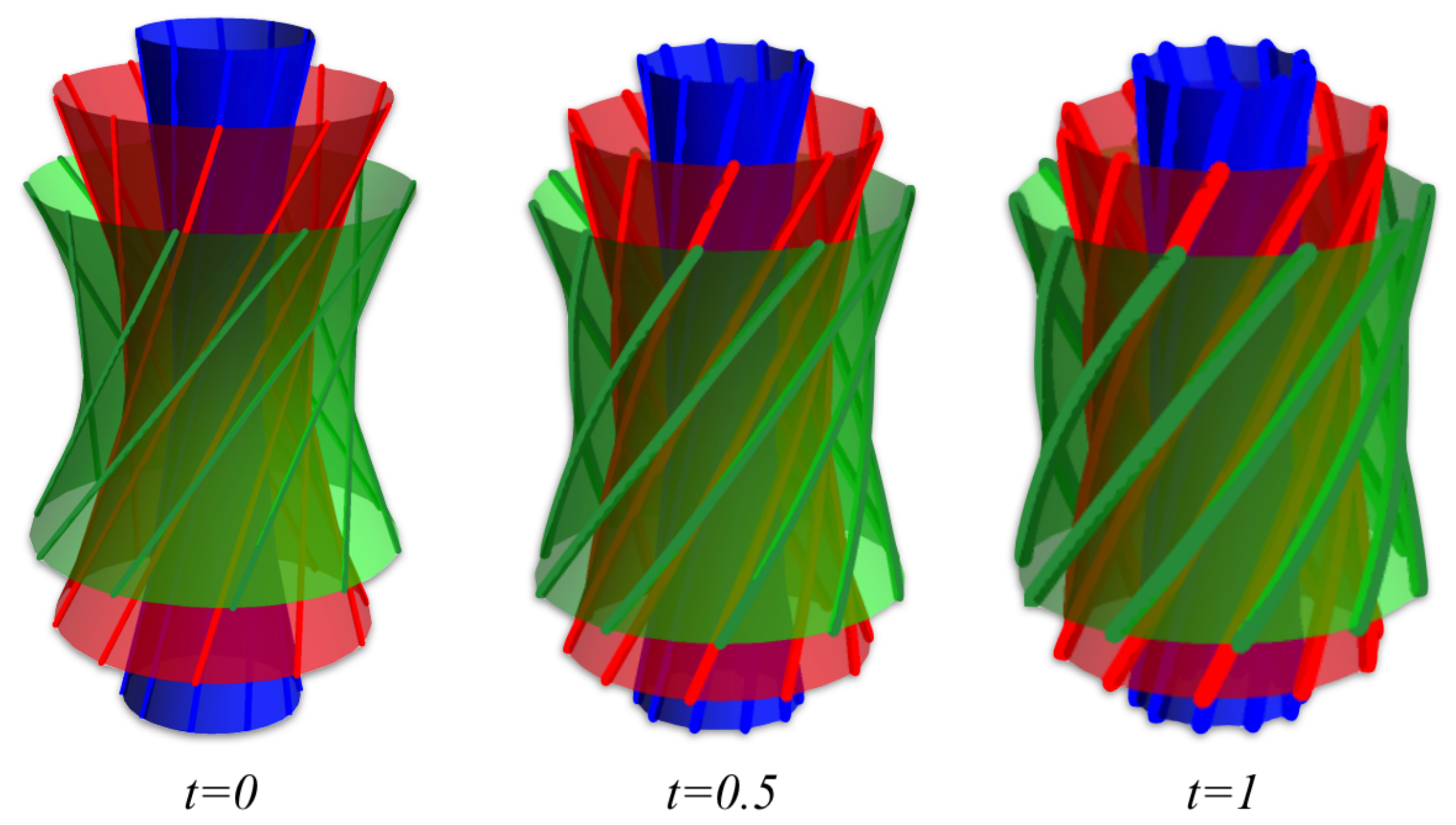}
\caption{Here we show the one-parameter family of projections that interpolates between the stereographic projection and the gnonomic projection, zoomed in near the origin. As $t$ changes from $0$ to $1$, the lines change from straight rulings on concentric hyperboloids to arcs of circles.}
\label{fig:hyper}
\end{figure}

\begin{figure}[h]
\centering
\includegraphics[width=\linewidth]{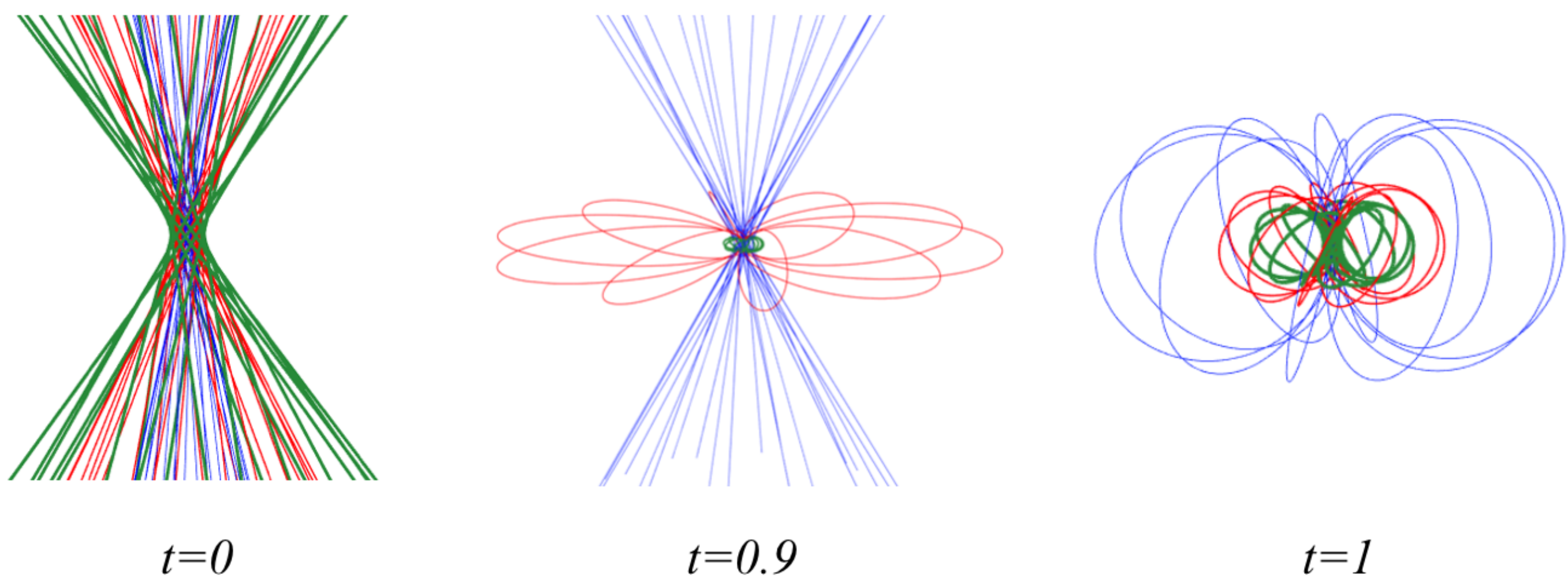}
\caption{Here we show the one-parameter family of projections that interpolates between the stereographic projection and the gnonomic projection, at large distances. As $t$ grows from $0$ to $1$, the projection changes from pure splay, to a mixed state where some lines curl around into circles while others fly off to infinity as pure splay, and finally to the more common linked-loop projection of the Hopf fibration.}
\label{fig:loops}
\end{figure}

To visualize this texture, it is instructive to consider a one-parameter family of projections that interpolates between the more familiar stereographic projection and the gnonomic projection.  Using the parameter $t\in[0,1]$ we have the projection from $\mathbb{S}^3$ to $\mathbb{R}^3$:
\begin{equation}
[x(t),y(t),z(t)] = [\tilde{x},\tilde{y},\tilde{z}]/(1+t-\tilde{w})
\end{equation}
where $t=0$ is the gnomonic projection and $t=1$ is the stereographic projection.  In Fig. \ref{fig:hyper} we zoom in near the origin to see the projected textures.  At $t=0$ the lines are straight and they form concentric hyperboloids.  As $t$ grows, the lines begin to curve until, at $t=1$ they become arcs of circles.  In Fig. \ref{fig:loops} we show the far field.  At large distances the $t=0$ projection becomes pure splay, while at $t=1$ we see the more common linked-loop projection of the Hopf fibration.  In between, at $t=0.9$ we can see a mixed state where some lines curl around into circles while others fly off to infinity as pure splay.

\begin{figure}[t]
\centering
  \includegraphics[width=\linewidth]{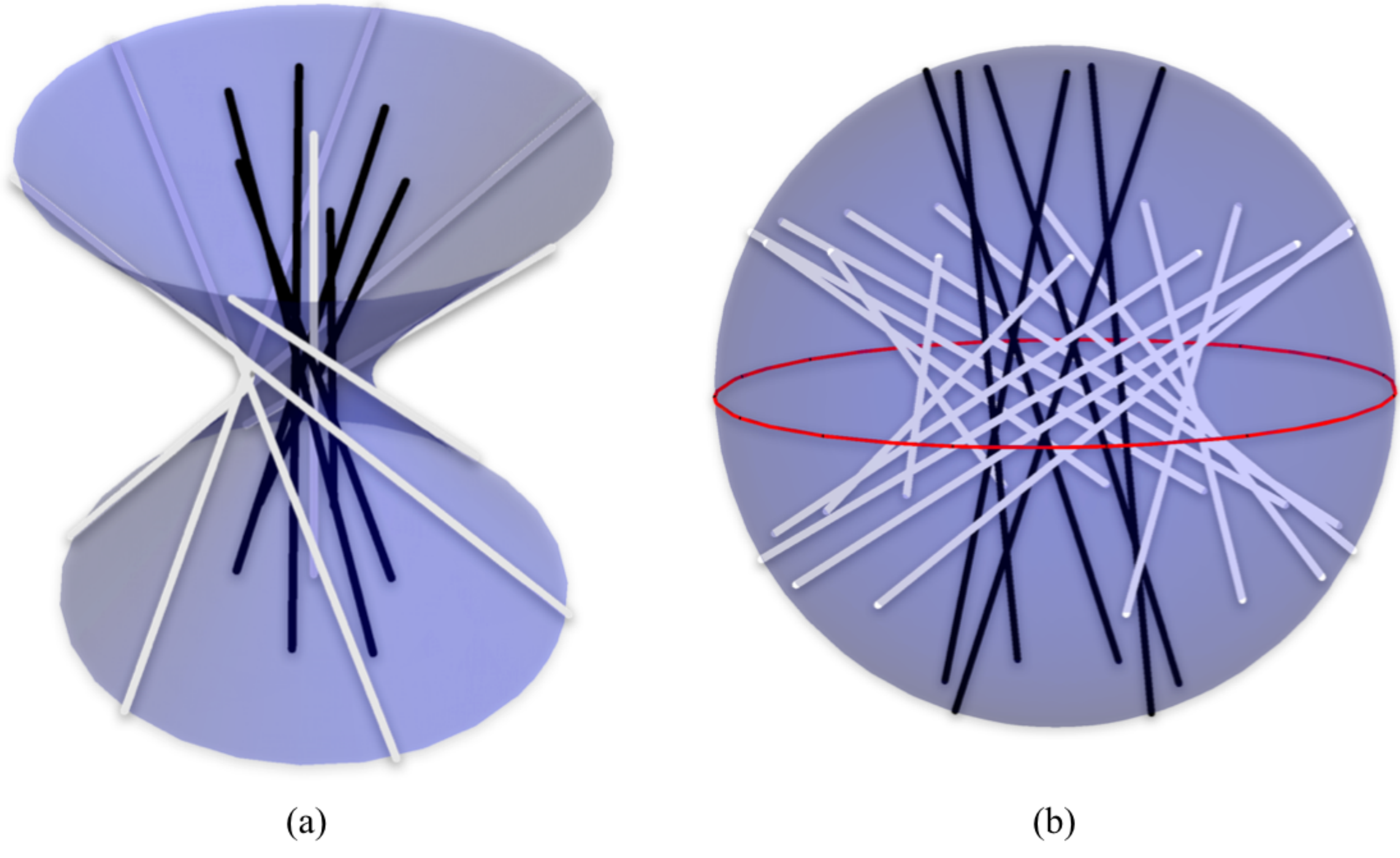}
\caption{In (a) we show a typical bundle shape. The bundle volume $M$ is defined by the hyperboloid generated by the outermost integral curves of the director.  The tops and bottoms are chosen as flat discs to seal the volume. In (b) we show a typical spherical droplet with the splay-twist structure inside. }
\label{fig:schematic}
\end{figure}

From \ref{nematicdirector}, we calculate the splay, twist, and bend of the Hopf projection and find
\begin{eqnarray}
\nabla \cdot \mathbf{n} &=& \frac{2z}{\sqrt{1+z^2}\sqrt{1+r^2}}\nonumber\\
\mathbf{n} \cdot \left( \nabla \times \mathbf{n} \right)&=&\left(\frac{1}{1+z^2} + \frac{1}{1+r^2}\right)\nonumber\\
\left({\bf n}\cdot\nabla\right){\bf n}&=&0
\end{eqnarray}
and we see that the bend vanishes, by construction.  In addition, the saddle-splay does not vanish,
\begin{align}
\nabla\!\cdot\!\left[{\bf n}\left(\nabla\!\cdot\!{\bf n}\right) - \left(\mathbf{n}\!\cdot\!\nabla\right){\bf n}\right]=2/({1+r^2}).
\end{align}

\section{Geometry and Stability}\label{geometry}

\begin{figure}[t]
\centering
\includegraphics[width=\linewidth]{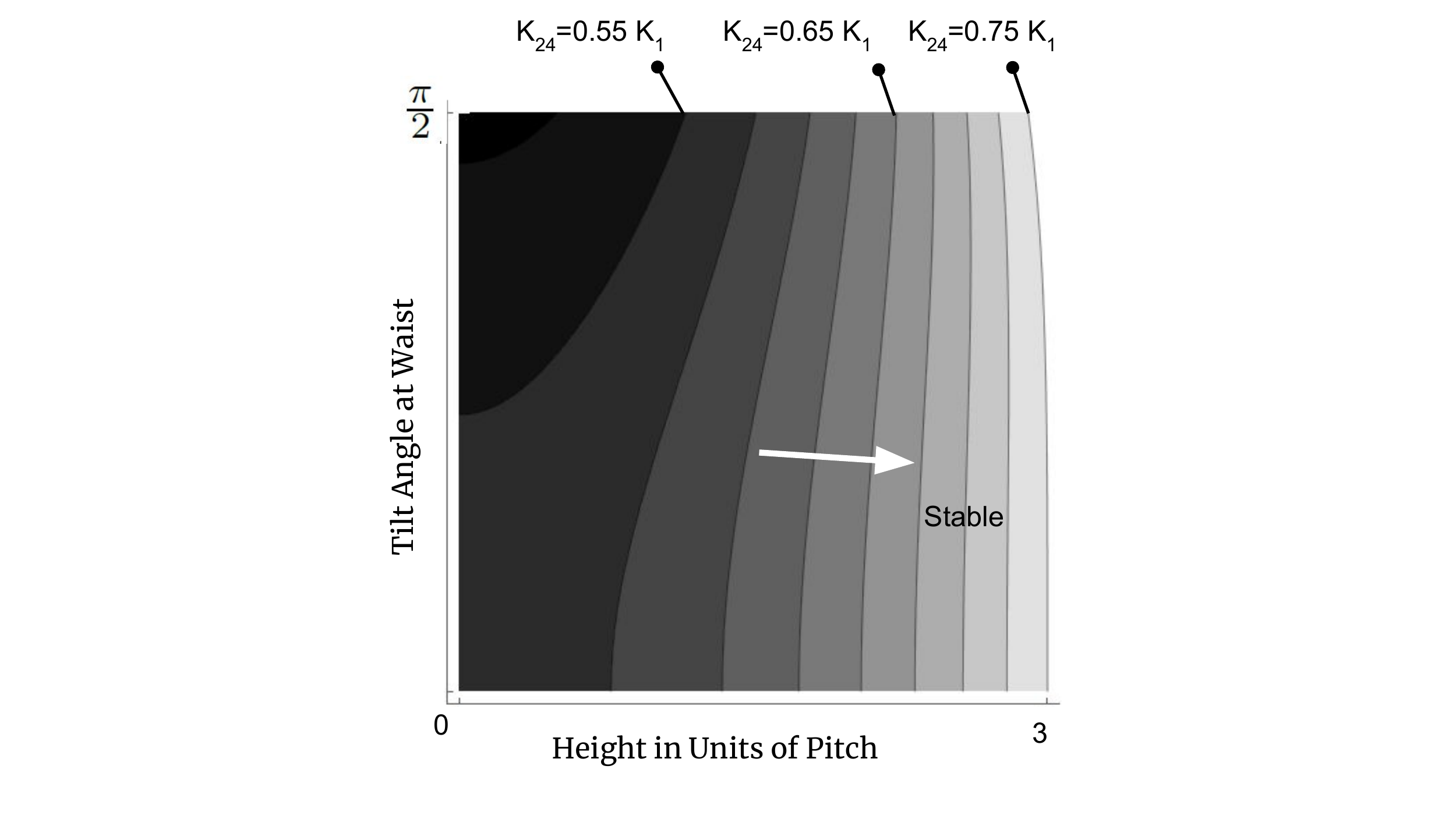}
\caption{The plot above is for the diabolo structure, with $K_1=K_2$ and $q=\frac{1}{2}$. We show the regions of stability as $K_{24}/K_1$ grows, with no anchoring.  As the shading gets lighter, the stability region grows up till the contour line for each value, including all the darker regions.}
\label{fig:diablo}
\end{figure}

Under what conditions is this texture stable?  Inspired by the ``diabolo'' textures found in \cite{livolant2000chiral}, we first consider a domain $M$ contained within a hyperboloid as depicted in Fig. \ref{fig:schematic}(a).  The director field is tangent to the hyperbola and intersects the discs at the top and the bottom in a swirl.   Compared to the cholesteric state, these diabolos are not stable when saddle-splay is negligible.  For a domain $M$, the Frank free energy is:
\begin{equation}
F=\int_{M} \!\!\!\!\!\!d^3\!x\left\{\frac{K_1}{2}\left[{\bf n}\left(\nabla\!\cdot\!{\bf n}\right)\right]^2 + \frac{K_2}{2}\left[{\bf n}\!\cdot\!\left(\nabla\!\times\!{\bf n}\right)-q\right]^2\right\} + \int_{\partial M} \!\!\!\!\!\!dA \left\{ W\left(\boldsymbol{\nu}\!\cdot\!\mathbf{n}\right)^2 - K_{24}\left(\boldsymbol{\nu}\!\cdot\!{\bf n}\right)\nabla\!\cdot\!{\bf n})\right\}
\label{eq:frank}
\end{equation}
where $\boldsymbol{\nu}$ is the outward pointing unit normal of $M$.  As usual, $K_1$, $K_2$, and $K_{24}$ are the splay, twist, and saddle-splay elastic constants while $W$ is the anchoring strength: negative $W$ favors homeotropic alignment.   We did not include the bend term as it vanishes identically.  The pitch $q$ allows for the possibility of a tendency to twist in the system.  In the following, we calculate the free energy for $K_1=K_2$, $W=0$, and $q=1/2$. The diabolo texture results in a stable domain as $K_{24}$ grows as depicted in Fig. \ref{fig:diablo}.  Larger values of $K_{24}/K_1$ lead to larger regions of stability (we have chosen the sign of the saddle-splay coupling so that positive values of $K_{24}$ favor this structure). Thus, for a diabolo system with saddle-splay and no explicit anchoring, the splay-twist may be more stable than a cholesteric.   In this particular geometry, the surface alignment $W$ leads to a complex contribution depending upon how the cholesteric phase sits in the diabolo and so, to demonstrate stability of this structure, we did not need to include it -- no new geometric insight would be gained by these complexities.

\begin{figure}[t]
\centering
\includegraphics[width=\linewidth]{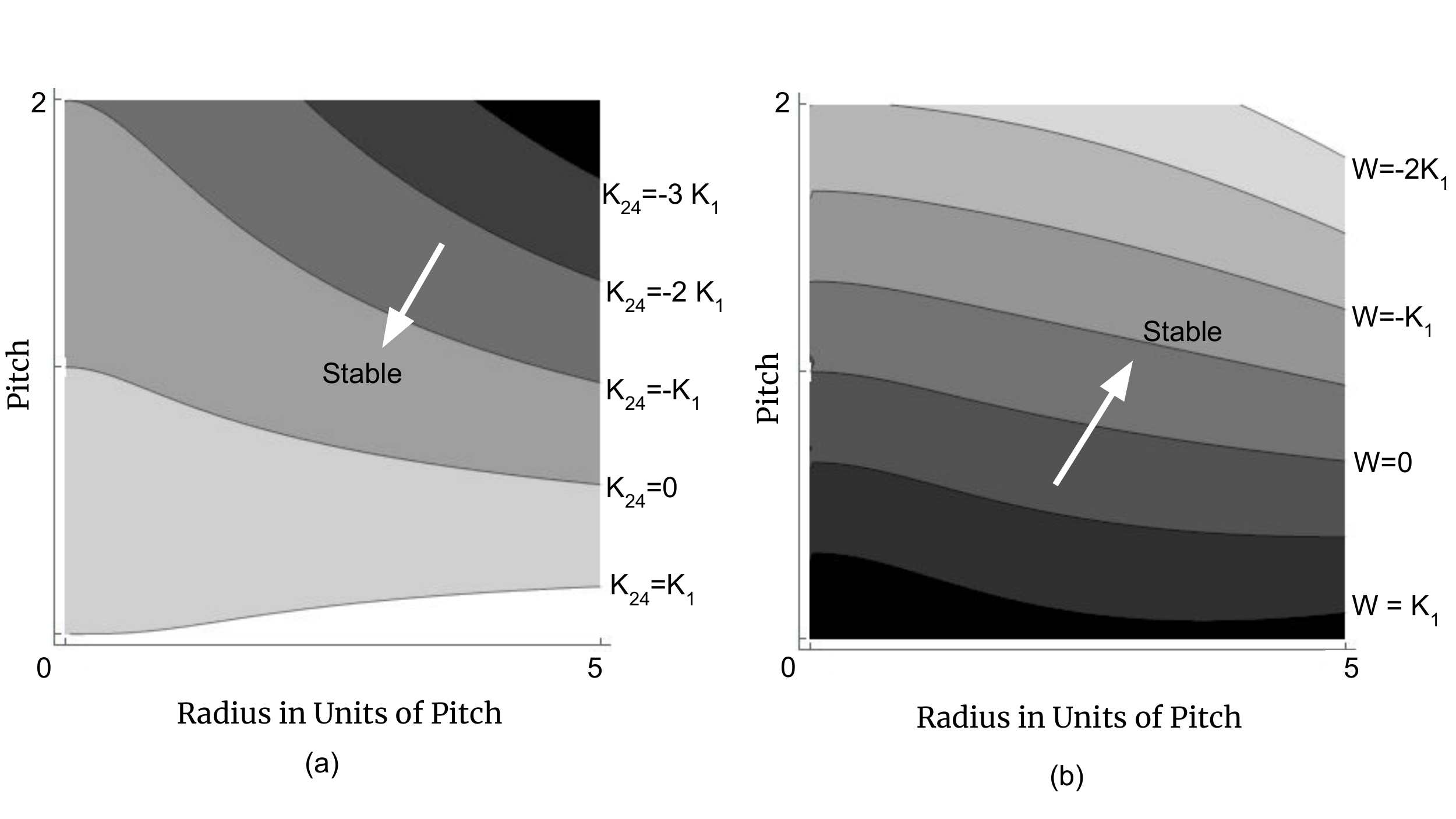}
\caption{Both the above plots are for the spherical droplet, with $K_1=K_2$. In (a) we show the regions of stability as $K_{24}/K_1$ grows, with no anchoring.  As the shading gets lighter, the stability region grows up till the contour line for each value, including all the darker regions. In (b) we show the regions of stability as $W/K_1$ becomes more negative for no saddle-splay, as compared with the free energy for a cholesteric of the same pitch. }
\label{fig:sphere}
\end{figure}

Next, we consider the possibility of isolating a splay-twist structure in a spherical droplet, as is shown in Fig. \ref{fig:schematic}(b). This is evocative of how hedgehog nematic defects have been stabilized in liquid crystalline droplets with homeotropic anchoring, although in our case, with added twist \cite{kleman2006topological}. The Frank free energy in (\ref{eq:frank}) is shown for a spherical geometry in Fig. \ref{fig:sphere} and now we allow $q$ to vary. In plot (a), as the saddle-splay, $K_{24}/K_1$, grows and changes from negative to positive, the region of stability, as compared to a cholesteric, grows. Even for negative saddle-splay a region of stability is seen for high values of the pitch. The plot in Fig. \ref{fig:sphere}(b) shows the regions of stability for non-zero anchoring and zero saddle-splay, as compared to a cholesteric. At low positive and small negative anchoring, the splay-twist is more stable than the cholesteric only for low values of the pitch $q$. As anchoring becomes more negative, and favors homeotropic anchoring more strongly, the splay-twist is stable even for large values of the pitch.

We have seen that the gnomonic projection of the Hopf fibration gives a particular arrangement of space filling hyperboloids.  However, the spacing of the hyperboloids can be distorted as long as no two helicoids intersect.  We embellish (\ref{spatialgnomonic}) to
\begin{align}
(x,y,z) = (\chi,\zeta,0) + \left[-\zeta,\chi,c\left(\sqrt{\chi^2+\zeta^2}\right)\right]s
\label{spatialgnomonic2}
\end{align}
where $c(\xi)>0$.
This representation generates a bend-free, twisted pattern with hyperboloids:
\begin{equation}
\frac{\rho^2}{\xi^2} -\frac{z^2}{c^2(\xi)} =1
\end{equation}
with $\xi^2=\chi^2+\zeta^2$, reducing to the gnomonic case when $c=1$ (or any constant).
What are the constraints on $c(\xi)$ so that the integral curves of the director field do not intersect?  Each hyperboloid is labeled by its waist radius $\xi_i$.  If two hyperboloids $\xi_0$ and $\xi_1>\xi_0$ intersect then they do so at the height $z$:
\begin{equation}
z^2=\frac{ \xi_1^2 - \xi_0^2}{\xi_0^2/c^2(\xi_0) - \xi_1^2/c^2(\xi_1)}
\end{equation}
A solution exists when $z^2\ge 0$ and so the hyperboloids avoid each other whenever 
\begin{equation}
\frac{c(\xi_0)}{c(\xi_1)}
 > \frac{\xi_0}{\xi_1}
\label{eq:inequality}
\end{equation}
Equivalently, $c(\xi)/\sqrt{c^2(\xi) + \xi^2}$ is a decreasing function -- the larger radii hyperboloids must tilt more and more toward the $xy$-plane.  As mentioned previously, this means that the characteristic curves from the $z=0$ plane never intersect.  In a finite geometry, it is possible that a virtual intersection could occur {\sl outside} the sample as discussed, for instance, in \cite{miller2019twist} in the context of viral rafts.  In that and other cases (\ref{eq:inequality}) would be modified by the sample size.  This also leads to the possibility of having a finite bundle such that at some radius $\bar\xi$ the $z$-component of the director field vanishes and thus the bundle stops.  We will cross those bridges if we come to them.  

Is the gnomonic projection special?  Consider concentric discs of the diabolo at $z=0$ of radii $\rho_1$ and $\rho_2$.  At height $z$ these discs are at new spacings, $\rho_i'$ with
\begin{equation}
\frac{\pi (\rho_1')^2-\pi\rho_1^2}{\pi\rho_1^2} = \frac{c^2(\rho_2)}{c^2(\rho_1)} \frac{\pi (\rho_2')^2-\pi\rho_2^2}{\pi\rho_2^2} 
\end{equation}
so that the isotropic expansion of each disk is not homogeneous unless $c(\rho_1)=c(\rho_2)$ -- precisely the gnomonic projection!  Thus, we see that among all projections, the gnomonic projection creates uniform expansion of a bundle of rigid lines.  In the liquid phase, the free energy is convex in the areal density of the lines (in the plane perpendicular to the $\hat z$-axis in this case) and thus a uniform density variation will minimise the free energy.

\begin{figure}[t]
\centering
\includegraphics[width=\linewidth]{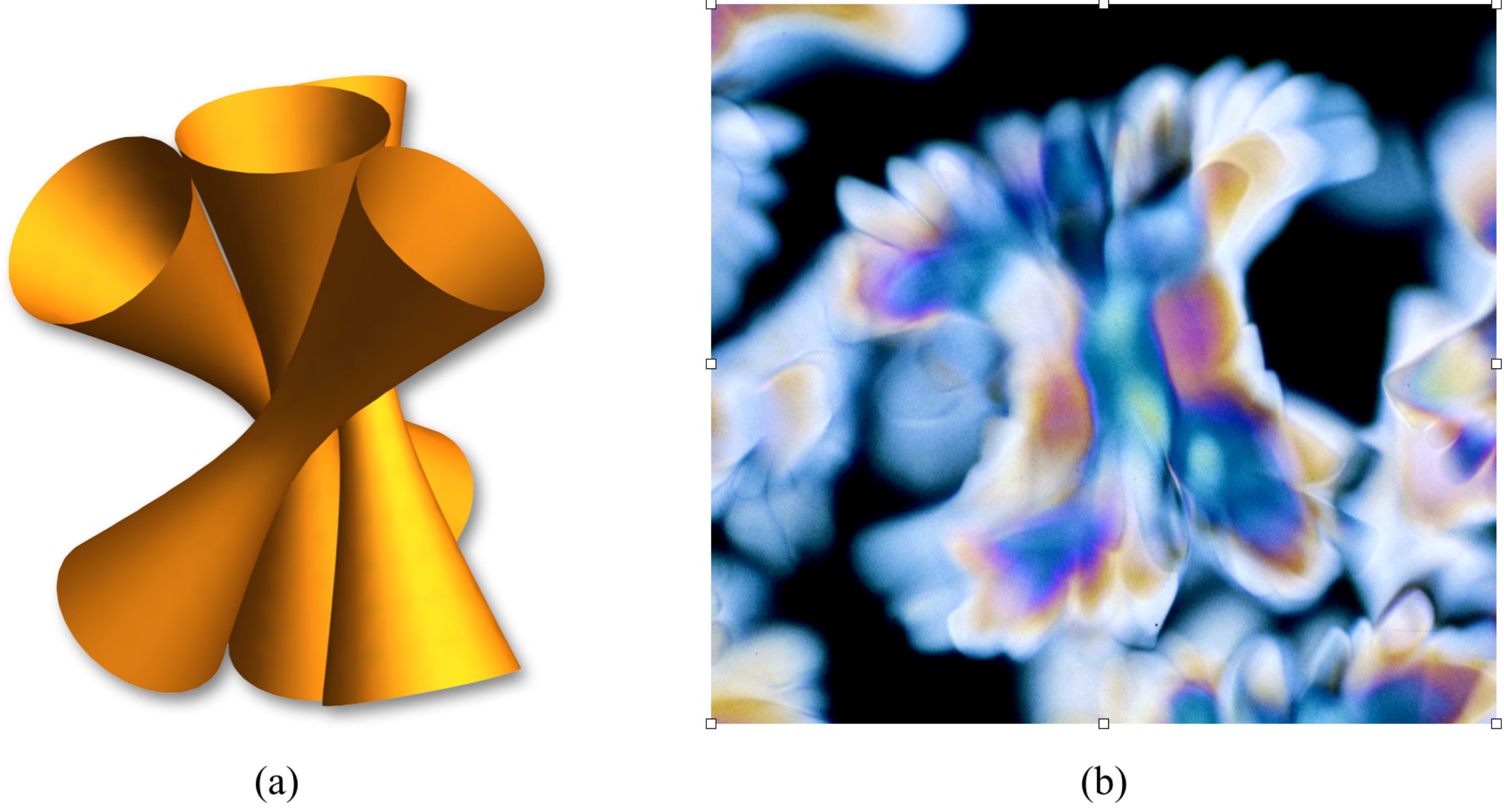}
\caption{In (a) we show a possible packing of the chiral diabolos and compare to (b) images of nucleosome core particles (previously unpublished image courtesy of A. Leforestier and F. Livolant \cite{livolant2000chiral}). }
\label{fig:exp}
\end{figure}

Each stable bundle might, in principle, act as a chiral constituent for a more complex arrangement as depicted in Fig. \ref{fig:exp}(a).  The chiral structure of the surface can lead to a chiral packing of each diabolo, something reminiscent of the observed packing of nucleosome core particles (NCPs) discovered by Leforestier and Livolant \cite{livolant2000chiral} shown in Fig. \ref{fig:exp}(b).  In that work, the NCPs formed a hexagonal columnar phase with two-dimensional crystalline order \cite{moire}. Because the plate-like stacking of the NCPs would enhance $K_3$ while doing little to the twist, we might expect straight, chiral distortions in these phases, as seen in other chromonic systems \cite{chromonic}. Incorporating hexagonal order into the bundle at the its waist, each successive $z$-slice is simply expanded uniformly by $\sqrt{1+z^2}$ with no shear, just a pure rotation.  The crystalline elasticity thus contributes
\begin{equation}
F_{\hbox{xtal}} \sim {B\over 2} \int_{-h}^h dz \pi\tan^2(\theta_0/2) z^2 = B{\pi\over 3}h^3\tan^2(\theta_0/2)
\end{equation}
where $B$ is the bulk modulus.  This would lead to stouter hyperboloids depending on the size of the twist penetration depth $\lambda=\sqrt{K_2/B}$.    

\section{Conclusions}
We have shown how the gnomonic projection of the Hopf fibration gives us a series of hyperboloids, the tangent vectors to which define a nematic structure with no bend. This structure can be the third in a series of new nematic phases defined by combinations of splay, twist and bend deformations, the twist-bend, splay-bend, and now the splay-twist. Further, the projection is one of a one parameter family of projections that goes from a stereographic projection that produces pure bend at one end, to the gnomonic projection at the other end. We have explored two different geometries in which such a phase could be stabilized -- a spherical droplet and a ``diabolo'' structure seen in experimental work on chiral discotic columnar germs formed by nucleosome particles\cite{livolant2000chiral}. Finally, we have shown that a variation of bend-free structures is possible, one of which is the gnomonic projection of the Hopf fibration which can provide a basis for bulk-energy-minimising bundles.  It would be interesting to extend this projection to other textures on $\mathbb{S}^3$ such as the Seifert fibrations considered in \cite{sadoc20093}.

\acknowledgments
We thank H. Ansell and E. Matsumoto for useful discussions and A. Leforestier and F. Livolant for providing us with their unpublished images.  This work was supported by NSF Grants DMR-1262047 and DMR-1720530.  This work
was supported by a Simons Investigator grant from the
Simons Foundation to R.D.K.



\begin{thebibliography}{99}

\bibitem{Frank}
Frank FC. 1958  I. Liquid crystals. On the theory of liquid crystals. {\em
  Discuss. Faraday Soc.} \textbf{25}, 19--28.

\bibitem{klemanbook}
Kleman M. 1983  Points, lines and walls. {\em Liquid Crystals, Magnetic Systems
  and Various Ordered Media. J. Wiley, New York}.

\bibitem{bouligand}
Bouligand Y, Denefle JP, Lechaire JP, Maillard M. 1985  Twisted architectures
  in cell-free assembled collagen gels: study of collagen substrates used for
  cultures. {\em Biology of the Cell} \textbf{54}, 143--162.

\bibitem{sadoc20093}
Sadoc J, Charvolin J. 2009  3-sphere fibrations: a tool for analyzing twisted
  materials in condensed matter. {\em Journal of Physics A: Mathematical and
  Theoretical} \textbf{42}, 465209.

\bibitem{pansu1987textures}
Pansu B, Dubois-Violette E. 1987  Textures of S3 blue phase. {\em Journal de
  Physique} \textbf{48}, 1861--1869.

\bibitem{dandoloff1993twisted}
Dandoloff R, Mosseri R. 1993  On the twisted director field of the blue phases.
  {\em EPL (Europhysics Letters)} \textbf{22}, 365.

\bibitem{sethna1983frustration}
Sethna JP. 1983  Frustration and curvature: Glasses and the cholesteric blue
  phase. {\em Physical review letters} \textbf{51}, 2198.

\bibitem{dandoloff1987blue}
Dandoloff R, Mosseri R. 1987  The blue phase: from S3 to a double-twisted tube
  in R3. {\em EPL (Europhysics Letters)} \textbf{3}, 1193.

\bibitem{mosseri2012hopf}
Mosseri R, Sadoc JF. 2012  Hopf fibrations and frustrated matter. {\em
  Structural chemistry} \textbf{23}, 1071--1078.

\bibitem{dozov2001spontaneous}
Dozov I. 2001  On the spontaneous symmetry breaking in the mesophases of
  achiral banana-shaped molecules. {\em EPL (Europhys. Lett.)} \textbf{56},
  247.

\bibitem{borshch2013nematic}
Borshch V, Kim YK, Xiang J, Gao M, J{\'a}kli A, Panov VP, Vij JK, Imrie CT,
  Tamba MG, Mehl GH et~al.. 2013  Nematic twist-bend phase with nanoscale
  modulation of molecular orientation. {\em Nature Commun.} \textbf{4}, 3635.

\bibitem{henderson2011methylene}
Henderson PA, Imrie CT. 2011  Methylene-linked liquid crystal dimers and the
  twist-bend nematic phase. {\em Liq. Cryst.} \textbf{38}, 1407--1414.
  
   
\bibitem{harden2006giant}
Harden J, Mbanga B, {\'E}ber N, Fodor-Csorba K, Sprunt S, Gleeson JT, Jakli A.
  2006  Giant flexoelectricity of bent-core nematic liquid crystals. {\em Phys.
  Rev. Lett.} \textbf{97}, 157802.

\bibitem{harden2010giant}
Harden J, Chambers M, Verduzco R, Luchette P, Gleeson JT, Sprunt S, J{\'a}kli
  A. 2010  Giant flexoelectricity in bent-core nematic liquid crystal
  elastomers. {\em Appl. Phys. Lett.} \textbf{96}, 102907.


\bibitem{mld}
Meyer, C, Luckhurst, G, Dozov, I. 2013. Flexoelectrically Driven Electroclinic Effect in the Twist-Bend Nematic Phase of Achiral Molecules with Bent Shapes. {\em Phys. Rev. Lett.} \textbf{111}, 067801.

\bibitem{shamid2013statistical}
Shamid SM, Dhakal S, Selinger JV. 2013  Statistical mechanics of bend
  flexoelectricity and the twist-bend phase in bent-core liquid crystals. {\em
  Phys. Rev. E} \textbf{87}, 052503.

\bibitem{prasang2008spontaneous}
Pr{\"a}sang C, Whitwood AC, Bruce DW. 2008  Spontaneous symmetry-breaking in
  halogen-bonded, bent-core liquid crystals: observation of a chemically driven
  Iso--N--N* phase sequence. {\em Chem. Commun.} \textbf{44}, 2137--2139.

\bibitem{meyer2015temperature}
Meyer C, Luckhurst G, Dozov I. 2015  The temperature dependence of the
  heliconical tilt angle in the twist-bend nematic phase of the odd dimer
  CB7CB. {\em J. of Materials Chem. C} \textbf{3}, 318--328.

\bibitem{cestari2011phase}
Cestari M, Diez-Berart S, Dunmur DA, Ferrarini A, de~la Fuente MR, Jackson DJB,
  Lopez DO, Luckhurst GR, Perez-Jubindo MA, Richardson RM, Salud J, Timimi BA,
  Zimmermann H. 2011  Phase behavior and properties of the liquid-crystal dimer
  1\ensuremath{'}\ensuremath{'},7\ensuremath{'}\ensuremath{'}-bis(4-cyanobiphenyl-4\ensuremath{'}-yl)
  heptane: A twist-bend nematic liquid crystal. {\em Phys. Rev. E} \textbf{84},
  031704.

\bibitem{chen2014twist}
Chen D, Nakata M, Shao R, Tuchband MR, Shuai M, Baumeister U, Weissflog W,
  Walba DM, Glaser MA, Maclennan JE et~al.. 2014  Twist-bend heliconical chiral
  nematic liquid crystal phase of an achiral rigid bent-core mesogen. {\em
  Phys. Rev. E} \textbf{89}, 022506.



\bibitem{mertelj2018splay} 
Mertelj A, Cmok L, Sebasti{\'a}n N, Mandle RJ, Parker RR, Whitwood AC, Goodby
  JW, {\v{C}}opi{\v{c}} M. 2018  Splay nematic phase. {\em Phys. Rev. X}
  \textbf{8}, 041025.

\bibitem{urbantke2003hopf}
Urbantke H. 2003  The Hopf fibration --- seven times in physics. {\em J. of
  geometry and phys.} \textbf{46}, 125--150.

\bibitem{chen2013generating}
Chen BGg, Ackerman PJ, Alexander GP, Kamien RD, Smalyukh II. 2013  Generating
  the Hopf fibration experimentally in nematic liquid crystals. {\em Physical
  review letters} \textbf{110}, 237801.

\bibitem{ng} 
Chamberlin, W.  1947.  {\em The Round Earth on Flat Paper, National Geographic, Washington D.C.}

\bibitem{CDG}
Callan CG, Dashen R, Gross DJ. 1978  Toward a theory of the strong
  interactions. {\em Phys. Rev. D} \textbf{17}, 2717--2763.
 
\bibitem{meyerrb}
Meyer, RB. 1982.  Macroscopic Phenomena in Liquid Crystal Polymers. In {\em Polymer Liquid Crystals, ed. by A. Ciferri, W.R. Kringbaum, and R.B. Meyer. Academic, New York}

\bibitem{selinger}
Duzgun A, Selinger JV, Saxena A. 2018  Comparing skyrmions and merons in chiral
  liquid crystals and magnets. {\em Phys. Rev. E} \textbf{97}, 062706.

\bibitem{livolant2000chiral}
Livolant F, Leforestier A. 2000  Chiral discotic columnar germs of nucleosome
  core particles. {\em Biophysical journal} \textbf{78}, 2716--2729.

\bibitem{kleman2006topological}
Kleman M, Lavrentovich OD. 2006  Topological point defects in nematic liquid
  crystals. {\em Phil. Mag.} \textbf{86}, 4117--4137.

\bibitem{miller2019twist}
Miller JM, Hall D, Robaszewski J, Sharma P, Hagan MF, Grason GM, Dogic Z. 2019
  All twist and no bend makes raft edges splay: Spontaneous curvature of domain
  edges in colloidal membranes. .

\bibitem{moire}
Kamien RD, Nelson DR. 1995  Iterated Moir\'e Maps and Braiding of Chiral
  Polymer Crystals. {\em Phys. Rev. Lett.} \textbf{74}, 2499--2502.
  
\bibitem{chromonic}
Tortuora, L, Lavrentovich, OD. 2001. Chiral symmetry breaking by spatial confinement in tactoidal droplets of lyotropic chromonic liquid crystals. {\em Proc. Natl. Acad. Sci.} \textbf{108}, 5163--5168.

\end{thebibliography}
\end{document}